\documentclass[11pt,a4paper]{article}
\usepackage{amsmath,amssymb}
\usepackage{cite}

\begin{document}

\title{
\begin{flushright}
{\normalsize Yaroslavl State University\\
             Preprint YARU-HE-06/07\\
             hep-ph/0611334} \\[20mm]
\end{flushright}
{\bf Comment on ``Spin light of neutrino in matter: \\a new type of 
electromagnetic radiation''}
}
\author{A.~V.~Kuznetsov$^a$\footnote{{\bf e-mail}: avkuzn@uniyar.ac.ru},
N.~V.~Mikheev$^{a}$\footnote{{\bf e-mail}: mikheev@uniyar.ac.ru}
\\
$^a$ \small{\em Yaroslavl State (P.G.~Demidov) University} \\
\small{\em Sovietskaya 14, 150000 Yaroslavl, Russian Federation}
}
\date{}

\maketitle

\begin{abstract}
We show that a criticism made in hep-ph/0610294 against our paper 
Mod. Phys. Lett. A21, 1769 (2006) has no ground. We confirm that all results 
of investigations of the so-called ``spin light of neutrino ($SL \nu$)'' are 
incorrect because of the medium influence on photon dispersion. With taking 
account of this influence, both the $SL \nu$ radiation rate and the total 
power are zero for neutrinos in all real astrophysical conditions. 
\end{abstract}


In an extended series of 
papers~\cite{LobStuPLB03,LobStuPLB04,DvoGriStuIJMP05,DvoStuJHEP02,StuNP05,
StuTerPLB05,GriStuTerGC05,LobPLB05,GriStuTerPLB05,LobDP05,StuJPA06}, 
there was proposed and 
investigated in detail the so-called ``spin light of neutrino'' ($SL \nu$), 
the process of the radiative neutrino transition 
$\nu_L \to \nu_R \gamma$, caused by the additional 
energy $W$ acquired in medium by the ultra-relativistic left-handed 
neutrino~\cite{Wolfenstein:1978}. In their analysis the authors neglected the 
medium influence on the photon 
dispersion. It looked unnatural, because the neutrino dispersion was defined by 
the weak interaction while the photon dispersion was defined by the 
electromagnetic interaction. In the paper~\cite{Kuznetsov:2006a} we evaluated 
the effective mass acquired by a photon in the astrophysical conditions used 
by those authors, and we have shown qualitatively that consideration of the 
radiative neutrino transition $\nu_L \to \nu_R \gamma$ without taking account 
of the photon dispersion in medium should be incorrect. In a short  
reply~\cite{Grigoriev:2006a}, the authors wrote that our analysis was 
wrong because ``the momentum conservation law has not been accounted for''. 
It was declared instead of performing an accurate application of this 
conservation law to the process. 

As we have shown further in the papers~\cite{Kuznetsov:2006b,Kuznetsov:2006c}, 
just the energy-momentum conservation law with the photon effective mass taken 
into account, has led to the threshold value for the initial neutrino energy:
\begin{equation}
E > E_0 \simeq \frac{m_\gamma^2}{2 \, W} \,, 
\label{eq:E_0}
\end{equation}
where $m_\gamma$ is the effective mass of the photon (plasmon) in medium. 
Evaluation of these threshold energies for different astrophysical 
situations shows that the value $E_0$ is always greater in orders of magnitude 
than the typical neutrino energy. Thus, the photon dispersion leaves no room for 
``spin light of neutrino'', except for a pure theoretical possibility when an 
ultra-high energy neutrino threads, e.g., a neutron star. The threshold neutrino 
energy for this case is $E_0 \simeq 10 \,{\rm TeV}$. 

In spite of a pure methodical meaning of this problem, we have performed in the 
paper~\cite{Kuznetsov:2006c} an accurate calculation of 
the process $\nu_L \to \nu_R \gamma$ width, as well as the energy and angular 
distributions of the photons created. 
We have evaluated the mean free path $L$ of an ultra-high energy 
neutrino with respect to the process, and we have obtained
\begin{equation}
L \gtrsim 10^{19} \, {\rm cm} \,, 
\label{eq:path}
\end{equation}
to be compared with the neutron star radius $\sim 10^{6}$ cm. 
This obviously illustrates the extreme weakness of the effect considered. 

In a recent preprint~\cite{Grigoriev:2006b} the authors have continued 
a discussion. They have totally ignored the basic point of our 
kinematical analysis, namely, the conclusion on the threshold 
behaviour of the process which closes the effect of ``spin light of neutrino'' 
in real astrophysical situations. 
The main objections by the authors~\cite{Grigoriev:2006b} to our 
paper~\cite{Kuznetsov:2006c} were fixed on our expression (18) for the 
amplitude squared, which was ill-defined, in their opinion. First, they made 
a statement that it was not positively-defined, and second, they 
asked a puzzling question, how this expression could be obtained at all. 
Now we show, that the amplitude squared was defined well enough, and that its 
positivity was unquestionable inside the kinematical region of the process.
We present also some details of calculation of the amplitude squared, which 
could be helpful for someone, while it should be a standard exercise for 
graduate students. 

At first sight, the positivity of the amplitude squared is not obvious:
\begin{equation}
|{\cal M}|^2 = 4 \, \mu_\nu^2 \, E^2 \left[ 2 \, W^2 \left(1 - \frac{\omega}{E} 
\right) - m_\gamma^2 \, \sin^2 \theta \right].
\label{eq:M^2}
\end{equation}
Here, $E = |\vec{p}|$ and $\omega$ are the neutrino and plasmon energies, 
$\theta$ is the angle between the 
initial neutrino momentum $\vec{p}$ and the plasmon momentum $\vec{k}$. 

Really, the plasmon mass $m_\gamma$ 
in the second negative term is much greater than the Wolfenstein energy $W$. 
However, one should wonder what is the restriction on the $\theta$ angle 
arising from the above-mentioned {energy-momentum conservation law}. 
Similarly, the differential probability for the classic weak process of muon 
decay contains the expression $(3 \, m_\mu - 4 \, E_e)$, however, nobody 
doubts about its positivity, knowing the range of the electron energies $E_e$. 

As for the ``problem'' of positivity of Eq.~(\ref{eq:M^2}), 
one should analyse the spatial distribution of final 
photons, and see that they are created inside the narrow cone 
with the opening angle $\theta_0$, see Eq. (21) of~\cite{Kuznetsov:2006c}:
\begin{equation}
\theta < \theta_0 \simeq \frac{\varepsilon - 1}{\varepsilon} \, 
\frac{W}{m_\gamma}\,,
\label{eq:theta}
\end{equation}
where $\varepsilon = E/E_0$, $E_0$ is the threshold neutrino 
energy~(\ref{eq:E_0}). 

There could be another hint for an attentive reader about the positivity 
of Eq.~(\ref{eq:M^2}). Integration in Eq. (19) of~\cite{Kuznetsov:2006c} was 
performed with the $\delta$ functions only, which means just the 
substitutions, and it led to Eq. (20). May be, it would be easier to check the 
positivity of the function $f (x, \varepsilon)$. Surely, a detailed 
quantitative analysis confirms these qualitative arguments.  

Now let us turn back to the calculation of the amplitude 
squared~(\ref{eq:M^2}). Let $p '$ and $q$ are the four-momenta of the final 
right-handed neutrino and photon, respectively, with $p '^2 = 0$ (we neglect 
the neutrino vacuum mass $m_\nu$ in our analysis), and $q^2 = m_\gamma^2$. 

The initial left-handed neutrino acquires in medium the additional 
energy $W$, and its four-momentum can be written as 
${\cal P}^\alpha = p^\alpha + W \, u^\alpha$, 
where ${\cal P}^\alpha$ is the neutrino four-momentum in medium, while 
$p^\alpha = (E,\, \vec{p})$ would form the neutrino four-momentum in vacuum, 
with $p^2 = 0$. The four-vector $u^\alpha$ of plasma velocity in its rest 
frame is $u^\alpha = (1,\, \vec{0})$. 

From the Lagrangian (17) of~\cite{Kuznetsov:2006c} one readily obtains the 
amplitude of the process $\nu_L \to \nu_R \gamma^{(\lambda)}$ with a 
creation of the plasmon with the polarization $\lambda$:
\begin{equation}
{\cal M}^{(\lambda)} = \mu_\nu \left({\bar u}_R ' \,\hat q 
\,{\hat \varepsilon}^{(\lambda)} 
\, u_L \right), 
\label{eq:M}
\end{equation}
where $u_L$ and $u_R '$ are the bispinor amplitudes for the initial left-handed 
and the final right-handed neutrinos. One should take care of the four-momentum 
which the bispinor amplitude $u_L$ depends on. To define it, one should write 
down the Dirac equation
\begin{equation}
\left({\hat{\cal P}} - \Sigma \right) u_L = 0\,, 
\label{eq:Dirac}
\end{equation}
where $\Sigma$ is the neutrino self-energy operator in plasma, 
$\Sigma = W \, \hat{u} \, (1 - \gamma_5)/2$. In the plasma rest frame one has 
$\Sigma = W \, \gamma_0 \, (1 - \gamma_5)/2$. Substituting the 
zero component ${\cal P}_0 = E + W$ and $\Sigma$ into Eq.~(\ref{eq:Dirac}), and
taking into account that $((1 - \gamma_5)/2) \, u_L = u_L$, one obtains:
\begin{equation}
\left(E \, \gamma_0 - \vec{p}\, \vec{\gamma} \right) u_L = 0\,. 
\label{eq:Dirac2}
\end{equation}
Thus, the bispinor amplitude $u_L$ depends on $p^\alpha = (E,\, \vec{p}), 
\; E = |\vec{p}|$.

The amplitude~(\ref{eq:M}) squared
\begin{equation}
\left|{\cal M}^{(\lambda)} \right|^2 = \mu_\nu^2 \, {\rm Tr} \, \left[\rho_L (p) \, 
{\hat \varepsilon}^{(\lambda)} \,\hat q \, \rho_R (p ') \,\hat q 
\,{\hat \varepsilon}^{(\lambda)} \right], 
\label{eq:M^2b}
\end{equation}
with the neutrino density martices substituted, $\rho_L (p) = u_L {\bar u}_L = 
\hat p \, (1 - \gamma_5)/2$, 
$\rho_R (p ') = u_R ' {\bar u}_R ' = \hat{p '} \, (1 + \gamma_5)/2$, is:
\begin{equation}
\left|{\cal M}^{(\lambda)} \right|^2 = 2 \, \mu_\nu^2 \, \left[ 2 (p q)\, (p ' q) - q^2 \, (p p ') 
- 2 \, q^2 \left( p \varepsilon^{(\lambda)}\right) \left( p ' \varepsilon^{(\lambda)}\right) 
\right]. 
\label{eq:M^2c}
\end{equation}
Using the energy-momentum conservation law and keeping in mind that 
$E > E_0 \gg W$, one obtains:
\begin{equation}
(p q) = W \,(E - \omega) + m_\gamma^2/2\,, \quad  
(p ' q) = W \,E - m_\gamma^2/2\,, \quad 
(p p ') = W \,\omega - m_\gamma^2/2\,. 
\label{eq:pq}
\end{equation}
Substituting Eqs.~(\ref{eq:pq}) into Eq.~(\ref{eq:M^2c}) and summarizing over 
the transversal plasmon polarizations:
\begin{equation}
\sum\limits_{\lambda} \left( p \varepsilon^{(\lambda)}\right) 
\left( p ' \varepsilon^{(\lambda)}\right) = E^2 \, \sin^2 \theta \,,
\qquad
\sum\limits_{\lambda} \left|{\cal M}^{(\lambda)} \right|^2 = \left|{\cal M} \right|^2\,,
\label{eq:sum}
\end{equation}
one readily obtains the amplitude squared~(\ref{eq:M^2}). 


As is seen from~\cite{Grigoriev:2006a,Grigoriev:2006b}, the authors believe 
that they have shown in Refs.~\cite{GriStuTerGC05,GriStuTerPLB05} that ``for 
the case of high-energy neutrinos 
the matter influence on the photon dispersion can be neglected'' because 
``plasma is transparent for electromagnetic radiation on frequencies greater 
than the plasmon frequency''. It is rather naive consideration. 
Really, one can see that from the side of kinematics the discussed process 
$\nu_L \to \nu_R \gamma$ in plasma is very similar to the process 
$\bar\nu_e e^- \to \tau^- \bar\nu_\tau$, where the high-energy electron 
anti-neutrino scattered off the electron in rest, creates 
the $\tau$ lepton. Neglecting the neutrino masses, one can see that the 
threshold value for the initial neutrino energy arises:
\begin{equation}
E > E_0 \simeq \frac{m_\tau^2}{2 \, m_e} \,, 
\label{eq:E_0_tau}
\end{equation}
to be compared with Eq.~(\ref{eq:E_0}). The similarity is deliberately 
not accidental. Both inequalities are caused by the minimal value of the 
Mandelstam $S$ variable 
which is equal to the mass squared of the heavy particle in the 
final state, $m_\tau^2$ (or $m_\gamma^2$ in our case). At the same time, 
the mass of the initial electron in rest is kinematically identical to 
the additional neutrino energy $W$. Taking the approach of the authors
~\cite{LobStuPLB03,LobStuPLB04,DvoGriStuIJMP05,DvoStuJHEP02,StuNP05,
StuTerPLB05,GriStuTerGC05,LobPLB05,GriStuTerPLB05,LobDP05,StuJPA06},
one should forget about the threshold~(\ref{eq:E_0_tau}) and conclude that 
the process $\bar\nu_e e^- \to \tau^- \bar\nu_\tau$
is open if only the medium (vacuum in this case) is transparent for 
$\tau$ leptons. 
 

It is interesting to note that it was not the first case when the plasma 
influence was taken into account for one participant of the physical process 
while it was not taken for other participant. The history is repeated. 

As E. Braaten wrote in Ref.~\cite{Braaten:1991}:

``In Ref.~\cite{Beaudet:1967}, it was argued that their calculation for the 
emissivities from photon and plasmon decay would break down at 
temperatures large enough that $m_\gamma > 2 \, m_e$, since the decay 
$\gamma \to e^+ e^-$ is then kinematically allowed. This statement, which 
has been repeated in subsequent 
papers~\cite{Dicus:1972,Munakata:1985,Schinder:1987,Itoh:1989}, is simply untrue. 
The plasma effects which generate the photon mass $m_\gamma$ also generate 
corrections to the electron mass such that the decay $\gamma \to e^+ e^-$ 
is always kinematically forbidden.''

Thus, the authors
~\cite{LobStuPLB03,LobStuPLB04,DvoGriStuIJMP05,DvoStuJHEP02,StuNP05,
StuTerPLB05,GriStuTerGC05,LobPLB05,GriStuTerPLB05,LobDP05,StuJPA06}
made the same mistake when they considered the plasma-induced additional 
neutrino energy $W$ and ignored the effective photon mass $m_\gamma$ arising 
by the same reason.  

In the recent papers~\cite{0611100,0611103,0611128} the authors have extended 
their approach to the so-called ``spin light of electron'' ($SL e$), 
$e_L \to e_R \gamma$. It should be mentioned that in these papers the authors 
have repeated just the same mistake of ignoring the photon dispersion 
in plasma. 

\section*{Acknowledgements}

We thank S.I. Blinnikov, M.I. Vysotsky, V.A. Novikov, L.B. Okun, and 
all the participants of the seminar at the Theoretical Department of 
ITEP (Moscow) for useful discussion. 

The work was supported in part 
by the Russian Foundation for Basic Research under the Grant No.~04-02-16253, 
and by the Council on Grants by the President of Russian Federation 
for the Support of Young Russian Scientists and Leading Scientific Schools of 
Russian Federation under the Grant No.~NSh-6376.2006.2.


\begin{thebibliography}{99}
%
%
\bibitem{LobStuPLB03}
A. Lobanov and A. Studenikin,
{\it Phys. Lett. B} {\bf 564 } 27 (2003).
%
\bibitem{LobStuPLB04}
A. Lobanov and A. Studenikin,
{\it Phys. Lett. B} {\bf 601} 171 (2004).
%
\bibitem{DvoGriStuIJMP05} 
M. Dvornikov, A. Grigoriev and A. Studenikin, 
{\it Int. J. Mod. Phys. D} {\bf 14} 309 (2005).
%
\bibitem{DvoStuJHEP02} 
M. Dvornikov and A. Studenikin,
{\it JHEP}, {\bf 09}, 016 (2002).
%
\bibitem{StuNP05} 
A. Studenikin, {\it Nucl. Phys. B (Proc. Suppl)}, {\bf 143} 570 (2005).
%
\bibitem{StuTerPLB05}
A. Studenikin and A. Ternov,  
{\it Phys. Lett. B} {\bf 608 } 107 (2005).
%
\bibitem{GriStuTerGC05} 
A. Grigoriev, A. Studenikin and A. Ternov, 
{\it Grav. \& Cosm.} {\bf 11}, 132 (2005). 
%
\bibitem{LobPLB05} 
A. E. Lobanov, 
{\it Phys. Lett. B} {\bf 619}, 136 (2005).
%
\bibitem{GriStuTerPLB05} 
A. Grigoriev, A. Studenikin and A. Ternov, 
{\it Phys. Lett. B} {\bf 622}, 199 (2005).
%
\bibitem{LobDP05} 
A. E. Lobanov, 
{\it Dokl. Phys.} {\bf 50}, 286 (2005). 
%
\bibitem{StuJPA06} 
A. Studenikin, 
{\it J. Phys. A: Math. Gen.} {\bf 39}, 6769 (2006).
%
%
\bibitem{Wolfenstein:1978}
L. Wolfenstein,
{\it Phys. Rev. D} {\bf 17}, 2369 (1978).
%
\bibitem{Kuznetsov:2006a} 
A. V. Kuznetsov and N. V. Mikheev, 
submitted to: Proceedings of the XL PNPI Winter School on Nuclear and Particle 
Physics and XII St.-Petersburg School on Theoretical Physics, St.-Petersburg, 
Repino, Russia, February 20-25, 2006; e-print hep-ph/0605114.
%
\bibitem{Grigoriev:2006a} 
A. Grigoriev, A. Lobanov, A. Studenikin and A. Ternov, %
E-print hep-ph/0606011, 2006.
%
\bibitem{Kuznetsov:2006b} 
A. V. Kuznetsov and N. V. Mikheev, 
submitted to: Proceedings of the XIV International Seminar "Quarks'2006", 
St.-Petersburg, Repino, Russia, May 19-25, 2006; e-print hep-ph/0606259.
%
\bibitem{Kuznetsov:2006c} 
A. V. Kuznetsov and N. V. Mikheev, 
{\it Mod. Phys. Lett. A} {\bf 21}, 1769 (2006).
%
\bibitem{Grigoriev:2006b} 
A. Grigoriev, A. Lobanov, A. Studenikin and A. Ternov, %
E-print hep-ph/0610294, 2006.
%
%
\bibitem{Braaten:1991}
E. Braaten, {\it Phys. Rev. Lett.}, {\bf 66}, 1655 (1991).
%
\bibitem{Beaudet:1967} 
G. Beaudet, V. Petrosian and E. E. Salpeter, 
{\it Astrophys. J.}, {\bf 150}, 979 (1967).
%
\bibitem{Dicus:1972}
D. A. Dicus, {\it Phys. Rev. D}, {\bf 6}, 941 (1972).
%
\bibitem{Munakata:1985}
H. Munakata, Y. Kohiyama and N. Itoh, 
{\it Astrophys. J.}, {\bf 296}, 197 (1985). 
%
\bibitem{Schinder:1987}
P. J. Schinder et al., 
{\it Astrophys. J.}, {\bf 313}, 531 (1987).
%
\bibitem{Itoh:1989}
N. Itoh et al., 
{\it Astrophys. J.}, {\bf 339}, 354 (1989).
%
%
\bibitem{0611100} 
A. Studenikin, E-print hep-ph/0611100, 2006.
%
\bibitem{0611103} 
A. Grigoriev, S. Shinkevich, A. Studenikin, A. Ternov and I. Trofimov,
E-print hep-ph/0611103, 2006.
%
\bibitem{0611128} 
A. Grigoriev, S. Shinkevich, A. Studenikin, A. Ternov and I. Trofimov,
E-print hep-ph/0611128, 2006.
%
\end{thebibliography}
\end{document}